\documentclass[preprint,review,12pt,sort&compress]{elsarticle}

\usepackage{graphics}
\usepackage{caption}
\usepackage{amssymb}
\usepackage{float}
\usepackage{hyperref}
\usepackage{geometry}
 \usepackage{layout}
\journal{Material Letters}

\begin{document}

\begin{frontmatter}



\title{Vacancy trapping behaviors of hydrogen atoms in Ti$_3$SiC$_2$: a first-principles study}


\author[label1]{Yi-Guo Xu}
\author[label2]{Xue-Dong Ou}
\author[label3]{Xi-Ming Rong\corref{label4}}

\cortext[label4]{Correspondence author. E-mail address: rongxm001@gmail.com. Tel: +86-21-65643967.
Fax: +86-21-65642787}

%\ead[label2]{Author to whom correspondence should be addressed.
%rongxm001@gmail.com.
%Tel: +86-21-65643967
%Fax: +86-21-65642787
%}

\address[label3]{Department of Optical Science and Engineering, Fudan University, Shanghai 200433, China}
\address[label1]{Institute of Modern Physics, Fudan University, Shanghai 200433, China}
\address[label2]{Department of Physics, Fudan University, Shanghai 200433, China}

\begin{abstract}
The behaviors of hydrogen (H) in MAX phase material Ti$_3$SiC$_2$ have been investigated using first-principles method. We show that a single H atom prefers to stay ~1.01 \AA\ down of the Si vacancy with solution energy of about -4.07 eV, lower than that in bulk Ti$_3$SiC$_2$. Multi H atoms exhibit a repulsive interaction at the Si vacancy. And up to five H atoms can be trapped by a Si vacancy without H$_2$ molecules formation. These results suggest the strong vacancy trapping
characteristic of H atoms in Ti$_3$SiC$_2$. Meanwhile, the barrier for H diffusion from an interstitial site to a vacancy is 1.17 eV, which is much larger than that in metals, indicating that to some extent H atoms can not easily migrate or aggregate to form bubble in Ti$_3$SiC$_2$.
\end{abstract}


\end{frontmatter}

% \linenumbers

\section{ Introduction}
\label{intro}
Ti$_3$SiC$_2$ is a typical MAX-phase (M$_{n+1}$AX$_n$, where n =1, 2 or 3, M is an early transition metal, A is an A-group element, and X is either carbon or nitrogen) material, which was firstly synthesized in 1967\cite{1967jeitschko329}. It has an unique combination of both metallic and ceramic properties, such as thermal-shock resistance\cite{2006zhang2401,1996barsoum1953JACS,1999raghy2855JACS},
high thermal conductivity\cite{1999barsoum429}, exceptional oxidation resistance\cite{2001barsoumc551}, significant ductility and high strength\cite{2000gilbert761,2001barsoum334}.
 This unique properties enable Ti$_3$SiC$_2$ a idea application under extreme condition, i,e, those proposed within future gas-cooled fast nuclear reactors (GFR) and further fusion programmes\cite{2010le766,2009nappe304JNM,2010liu149JNM,2010whittle4362AM,
2011nappe1503,2011nappe53JNM}. Under neutrons irradiation in reactors, large amounts of vacancy defects and hydrogen impurities would be produced continually in the structural materials.
It is well known that the behavior of hydrogen plays a very important role in mechanical properties of metals. H atoms has been shown to assist the vacancy formation and also can be trapped by the vacancy\cite{1993condon1JNM,1989nordlander1990PRB}. The H atom has low energy barrier in metals, and usually diffuse, segregate, form bubble and finally degrade the mechanical properties of materials\cite{2007arkhipov1168JNM}.
Previous researches show that the unique properties of Ti$_3$SiC$_2$ were contributed to the nanolaminate crystal structure and the bonding properties\cite{2001zhou10001}.
And the structure and atomic bonding properties of Ti$_3$SiC$_2$ are totally different from that in metals.
Therefore, maybe H atoms in Ti$_3$SiC$_2$ behave very differently.
 To date, most of theoretical studies focused on the chemical bonding of defect-free Ti$_3$SiC$_2$,
 Litter work has been done on the effect of H impurity. The behavior of H atoms in Ti$_3$SiC$_2$ with vacancies is not fully understood. Therefore, to understand the interaction between H atom and vacancies in Ti$_3$SiC$_2$ is in great need.

In the present work, we investigate behaviors of H atoms in Ti$_3$SiC$_2$ using first-principles calculation. The stability, vacancy trapping and diffusion behaviors of H in Ti$_3$SiC$_2$ were all discussed. These computational results might be not only beneficial for designing and processing of Ti$_3$SiC$_2$-based materials in realistic applications, but also beneficial for understanding the physical essence.


 \section{Method}

 The first-principles calculations of the molecular structure were performed using the density functional theory (DFT) and the pseudopotential plane-wave method implemented in the VASP code\cite{1993kresse558PRB,1996kressee11169PRB}. A gradient-corrected form of the exchange correlation functional generalized gradient approximation (GGA-PW91) was used\cite{1994blochl17953PRB,1992perdew6671PRB}.The calculations were made using the plane-wave cutoff energy of 500 eV. And a 2$\times$1$\times$1 and a 2$\times$2$\times$1 super cell were performed. The $k$-points sampling Brillouin zone were 9$\times$9$\times$4 generated by the Monkhorst-Pack scheme\cite{1977chadi1746PRB}. The energy minimization was converged at atomic forces less than 0.001 eV/\AA  .

  \section{Results and discussion}
 \subsection{H atom in bulk Ti$_3$SiC$_2$}
 Ti$_3$SiC$_2$ has a hexagonal crystal structure with a space group of P63/$mmc$, where Ti occupies $2a$ and $4f$, Si $2b$, and C $4f$ Wyckoff position, as shown in Fig. \ref{fig1}. To distinguish the Ti atoms in two different coordination environments, we denote Ti atoms sitting at $2a$ sites as Ti (1), and Ti in 4f as Ti (2). After fully relaxing the structure of Ti$_3$SiC$_2$, we obtained the lattice constants of a=3.075 \AA\ and c=17.561 \AA, in good agreement with the previous work\cite{1999sun1441PRB}.

In order to find the most stable site of a single H atom in perfect Ti$_3$SiC$_2$, we calculated the solution energy of possible sites for a H atom. The solution energy of interstitial for a H atom in the Ti$_3$SiC$_2$ is defined as
       \begin{equation}\label{equ1}
        E^s(H)=E(ref+H)-E(ref)-E_H,
        \end{equation}
where $E(ref+H)$ is the energy of Ti$_3$SiC$_2$ with a single H atom, $E(ref)$ is the energy of perfect Ti$_3$SiC$_2$ crystal, $E_H$ is the energy of isolated H atom. There are three possible interstitial positions with large free volumes in Ti$_3$SiC$_2$ (Fig. \ref{fig1}): a tetrahedral interstitial (I-Ti) surrounded by three Ti(1) and one Ti(2) atoms, a hexahedral interstitial (I-SiTi) surrounded by three Si and two Ti(2) atoms, and a tetrahedral interstitial (I-SiC) surrounded by three Si and one C atoms. The solution energy is calculated to be -2.215 eV, -2.870 eV and -2.894 eV for I-Ti, I-SiTi and I-SiC, respectively, indicating I-SiC is the most stable position for a single H in perfect Ti$_3$SiC$_2$.

In order to investigate the interaction between two H atoms in buck Ti$_3$SiC$_2$, we firstly placed two H atoms into I-Ti, I-SiTi and I-SiC three sits, and then calculated the binding energy between two H atoms, which is defined as
        \begin{equation}\label{equ2}
            E^s(H)=E(ref+2H)+E(ref)-2E(ref+H),
        \end{equation}
by such definition, negative binding energy indicates attraction, while positive one indicates repulsion. The binding energy for two H atoms in I-SiC is 0.738 eV, indicating a repulsive H-H interaction, and the corresponding distance of the two H atoms is 1.53 \AA. In I-SiTi the binding energy is 0.008 eV, exhibiting a week repulsive interaction between two H atoms. and the distance is 2.11 \AA. Similarly, in I-Ti the two H atoms exhibit a repulsive interaction with binding energy of 0.500 eV, and the distance between the two H atoms is 1.57 \AA. The results show that two H atoms are repulsive in most cases in perfect Ti$_3$SiC$_2$. Additionally, with the distance increasing, the repulsive interaction gets weaker. When the distance becomes 2.11 \AA, the binding energy drops to only 0.008 eV. So H atom cannot be self-trapped. The distance of two H atoms is much larger than that in a H$_2$ molecule (0.76 \AA), indicating that the H$_2$ molecule cannot be formed directly.
\subsection{H trapping at a Si vacancy }
In metals, H atoms are seen to be trapped by the vacancy in theoretical predictions and experimental measurements\cite{1982Norskov1420PRL,1985besenbacher852PRL,1989nordlander1990PRB}. Vacancy reduces charge density in its vicinity to provide an isosurface for collective H binding in W, causing H segregation and hence H bubble nucleation in vacancy surface. First-principles calculations suggested that to form a Si vacancy in Ti$_3$SiC$_2$ is easier than other vacancies\cite{2012jia23ML}. For this reason, we then introduced a Si vacancy in Ti$_3$SiC$_2$ with a supercell size of 2$\times$2$\times$1.

To determine the number of H atom that a vacancy can accommodate, we calculate the trapping energy of additional H atoms segregating to the vacancy, which is defined as
  \begin{equation}\label{equ3}
  %\scriptsize
            E_{trap}=E(nH,V+ref)-E((n-1)H,V+ref)-E(H+ref)+E(ref),
        \end{equation}
where  $E(nH,V+ref)$ is the energy of Ti$_3$SiC$_2$ with $nH$ atoms and a single vacancy. A negative value for the trapping energy represents the energy gain when the H atoms are trapped at a single vacancy site relative to being dispersed at a interstitial site. The trapping energy  as a function of the numbers of H atoms in a Si vacancy is illustrated in Fig. \ref{fig2}.

By our calculation, we found that H atom occupation surrounding the vacancy in Ti$_3$SiC$_2$ satisfies the rule of 'optimal charge density'\cite{2009liu172103PRB,2010zhou115010NF,2010zhoui025016NF,2012zhang216JNM,20136alkhamees6JNM,2013Sun395JNM}, as shown in Fig. \ref{fig3}. When a single H atom was put at a Si vacancy, it prefers to stay ~1.01 \AA\ down of the Si vacancy rather than a Si substitution site. The trapping energy of a single H atom to a Si vacancy is -1.101 eV, which is negative. And the solution energy of H in a Si vacancy is -4.065 eV, which is much lower than that of H atom in perfect Ti$_3$SiC$_2$, indicting that vacancies favorably trap H atoms. It can be understood by an early model of homogeneous electron gas\cite{1981puska3037PRB,1985besenbacher55,1987norskov475}. H solution energy decreases with the electron density decreases. It means that the H atom has lower solution energy at where the electron density is reduced. The isovalue of electron density surrounding the vacancy is calculated to be about 0.06 e/\AA$^3$ [Fig. \ref{fig3}(a)], lower than the value of 0.13 e/\AA$^3$ for H surrounding I-SiC site in prefect Ti$_3$SiC$_2$ (Fig. \ref{fig4}). Therefore, the presence of vacancy reduces the electron density, and thus results in lower H solution energy in the vacancy. So the vacancy favorably trap H atoms.

For two H atoms, the trapping energy is -1.362 eV. They formed a dumbbell with H atoms residing on the up and down sides of a Si vacancy Fig. \ref{fig3}(c). The distance of the two H atoms is about 2.25 \AA, which is much larger than that in a H$_2$ molecule (0.76 \AA), So two H atoms cannot directly form a H$_2$ molecule. But the binding energy of two H atoms in a Si vacancy is negative, -0.262 eV, exhibiting an attractive H-H interaction.
When adding the third H atom in the Si vacancy, the trapping energy becomes -0.214 eV. And the third H atom will stay ~2.02 \AA\ up side of Ti(2) atom. For the fourth and fifth H atom, the trapping energy is -0.202 eV and -0.263 eV, respectively. Similarly, they both stay ~2.02 \AA\ up side of Ti(2) atom symmetrically. And the third, fourth, and fifth H atom form a regular triangle with a H atom residing at each vertex. The side length of the triangle is 3.52 \AA. It indicates that H$_2$ molecule can not be formed. But For the sixth H atoms, the tapping energy becomes positive, 0.184 eV, suggesting that the sixth H is energetically unfavorable been trapped by a Si vacancy. Therefore, the maximal number of H atoms that can be trapped by a Si vacancy is five.

\subsection{Diffusion of H in Ti$_3$SiC$_2$}
We also calculated the diffusion energy barrier of H in Ti$_3$SiC$_2$ with and without a Si vacancy. Here we use a drag method at a fixed volume and constrain the atomic positions to relax in a hyperplane perpendicular to the vector from the initial to finial positions\cite{2004fu175503PRL}.
As discussed above, in buck Ti$_3$SiC$_2$, I-SiC is the most stable site for a H atom. Therefore, as illustrated in Fig. \ref{fig5}(a), the H atom in I-SiC (Site 1) can migrate to its equivalent site (Site 3) or its nearest neighboring I-SiC site (Site 2). Then
we calculated the diffusion barrier for the optimal diffusion path from I-SiC to its equivalent site   (1-3 path) and the one from I-SiC to its nearest neighboring I-SiC (1-2 path). The diffusion barrier along 1-3 path is calculated to be 0.69 eV and along 1-2 path is calculated to be 2.64 eV.
When introducing a Si vacancy in Ti$_3$SiC$_2$ [Fig. \ref{fig5}(b)], a H atom in the stable site (Site 1) around a Si vacancy can migrate to a I-SiC site (Site 3 or Site 4) (Site 3 and Site 4 are equivalent). So we calculated the barrier from Site 1 to Site 4. H atom has two possible paths of diffusion: path 1 is from Site 1 to Sit 4 in a straight line (1-4 path); path 2 is from Site 1, then passing by Site 2, finally to the Site 4 (1-2-4 path). The diffusion barrier along 1-2-4 path is calculated to be 2.83 eV, lower than the value along 1-4 path, as shown in Fig 6. The saddle point of path 2 is just at Site 2. Therefore, the most possible path is 1-2-4 path. And Sit 1 is more stable than Sit 4 by 1.66 eV, indicating that the vacancy behaves like a trap for H atoms. And the barrier for a H atom diffusion from a I-SiC site to a vacancy is 1.17 eV, larger than that in metals\cite{2009liu172103PRB,2009xu3170,2010duan109JNM,2009ismer184110PRB}. The results imply that H atom is not easy to migrate in $_3$SiC$_2$, even when a vacancy exists. Therefore, comparing with metals, H atoms to some extent can not easily migrate, segregate to form bubble.

\section{Conclusion}
We investigated the behaviors of H atoms in MAX phase material using a first-principles method. We found that H occupation surrounding the vacancy in Ti$_3$SiC$_2$ satisfies the rule of ¡®optimal charge density¡¯. The preferential site for single H is not Si vacancy center but 1.01 \AA\ down of the Si vacancy with solution energy of about -4.07 eV, which is much lower than that in a perfect Ti$_3$SiC$_2$. And two H atoms exhibit an attractive interaction at the vacancy with a binding energy of about -0.26 eV, which is different from that in a perfect Ti$_3$SiC$_2$. Two H atoms exhibit a repulsive interaction. According to the trap energy, a Si vacancy can hold up to five H atoms, but no H$_2$ molecule was formed. These results suggest the strong vacancy trapping characteristic of H atoms in Ti$_3$SiC$_2$. However, H migration in Ti$_3$SiC$_2$ is not easy, even when a vacancy exists. The diffusion barrier for H atom from a I-SiC site to its nearest neighbouring I-SiC site in a straight line is 2.64 eV in perfect Ti$_3$SiC$_2$. And the barrier for H diffusion from an interstitial site to a vacancy is 1.17 eV, larger than that in metals, suggesting that H atoms to some extent can not easily migrate and aggregate to form bubble. These results provide a useful reference to understand the behaviors of H atoms in Ti$_3$SiC$_2$.

%%insert equation
%\begin{equation}\label{equ4}
 % v_i ^{new}=(1-\theta)^{1/2} v_i ^{old}+ \theta^{1/2} v_i ^T (\xi) ,(i=x,y,z)
%\end{equation}

%% insert figure
% \begin{figure}
% \end{figure}
%\ref{}

%% References
\newpage
\noindent{\large{References}}

\footnotesize
\setlength{\bibsep}{0ex}  % vertical spacing between references

\bibliographystyle{model3-num-names}

%\bibliography{jnm}

\begin{thebibliography}{40}
\expandafter\ifx\csname natexlab\endcsname\relax\def\natexlab#1{#1}\fi
\providecommand{\bibinfo}[2]{#2}
\ifx\xfnm\relax \def\xfnm[#1]{\unskip,\space#1}\fi
%Type = Article
\bibitem[{Jeitschko and Nowotny(1967)}]{1967jeitschko329}
\bibinfo{author}{W.~Jeitschko}, \bibinfo{author}{H.~Nowotny},
  \bibinfo{journal}{Monatsh. Chem.} \bibinfo{volume}{98} (\bibinfo{year}{1967})
  \bibinfo{pages}{329--337}.
%Type = Article
\bibitem[{Zhang et~al.(2006)Zhang, Zhou, Bao, and Li}]{2006zhang2401}
\bibinfo{author}{H.~Zhang}, \bibinfo{author}{Y.~Zhou},
  \bibinfo{author}{Y.~Bao}, \bibinfo{author}{M.~Li}, \bibinfo{journal}{J.
  Mater. Res.} \bibinfo{volume}{21} (\bibinfo{year}{2006})
  \bibinfo{pages}{2401--2407}.
%Type = Article
\bibitem[{Barsoum and El~Raghy(1996)}]{1996barsoum1953JACS}
\bibinfo{author}{M.~W. Barsoum}, \bibinfo{author}{T.~El~Raghy},
  \bibinfo{journal}{J. Am. Ceram. Soc.} \bibinfo{volume}{79}
  (\bibinfo{year}{1996}) \bibinfo{pages}{1953--1956}.
%Type = Article
\bibitem[{El-Raghy et~al.(1999)El-Raghy, Barsoum, Zavaliangos, and
  Kalidindi}]{1999raghy2855JACS}
\bibinfo{author}{T.~El-Raghy}, \bibinfo{author}{M.~W. Barsoum},
  \bibinfo{author}{A.~Zavaliangos}, \bibinfo{author}{S.~R. Kalidindi},
  \bibinfo{journal}{J. Am. Ceram. Soc.} \bibinfo{volume}{82}
  (\bibinfo{year}{1999}) \bibinfo{pages}{2855--2860}.
%Type = Article
\bibitem[{Barsoum et~al.(1999)Barsoum, El-Raghy, Rawn, Porter, Wang, Payzant,
  and Hubbard}]{1999barsoum429}
\bibinfo{author}{M.~Barsoum}, \bibinfo{author}{T.~El-Raghy},
  \bibinfo{author}{C.~Rawn}, \bibinfo{author}{W.~Porter},
  \bibinfo{author}{H.~Wang}, \bibinfo{author}{E.~Payzant},
  \bibinfo{author}{C.~Hubbard}, \bibinfo{journal}{J. Phys. Chem. Solids}
  \bibinfo{volume}{60} (\bibinfo{year}{1999}) \bibinfo{pages}{429--439}.
%Type = Article
\bibitem[{Barsoum et~al.(2001)Barsoum, Tzenov, Procopio, El-Raghy, and
  Ali}]{2001barsoumc551}
\bibinfo{author}{M.~Barsoum}, \bibinfo{author}{N.~Tzenov},
  \bibinfo{author}{A.~Procopio}, \bibinfo{author}{T.~El-Raghy},
  \bibinfo{author}{M.~Ali}, \bibinfo{journal}{J. Electrochem. Soc.}
  \bibinfo{volume}{148} (\bibinfo{year}{2001}) \bibinfo{pages}{C551--C562}.
%Type = Article
\bibitem[{C.J.~Gilbert and Ritchie(2000)}]{2000gilbert761}
\bibinfo{author}{M.~B. T. E.-R. A.~T. C.J.~Gilbert, D.R.~Bloyer},
  \bibinfo{author}{R.~Ritchie}, \bibinfo{journal}{Scripta mater.}
  \bibinfo{volume}{42} (\bibinfo{year}{2000}) \bibinfo{pages}{761¨C767}.
%Type = Article
\bibitem[{Barsoum and El-Raghy(2001)}]{2001barsoum334}
\bibinfo{author}{M.~W. Barsoum}, \bibinfo{author}{T.~El-Raghy},
  \bibinfo{journal}{Am. Scientist} \bibinfo{volume}{89} (\bibinfo{year}{2001})
  \bibinfo{pages}{334--343}.
%Type = Article
\bibitem[{Le~Flem et~al.(2010)Le~Flem, Liu, Doriot, Cozzika, and
  Monnet}]{2010le766}
\bibinfo{author}{M.~Le~Flem}, \bibinfo{author}{X.~Liu},
  \bibinfo{author}{S.~Doriot}, \bibinfo{author}{T.~Cozzika},
  \bibinfo{author}{I.~Monnet}, \bibinfo{journal}{Int. J. Appl. Ceram. Techn.}
  \bibinfo{volume}{7} (\bibinfo{year}{2010}) \bibinfo{pages}{766--775}.
%Type = Article
\bibitem[{Napp{\'e} et~al.(2009)Napp{\'e}, Grosseau, Audubert, Guilhot, Beauvy,
  Benabdesselam, and Monnet}]{2009nappe304JNM}
\bibinfo{author}{J.-C. Napp{\'e}}, \bibinfo{author}{P.~Grosseau},
  \bibinfo{author}{F.~Audubert}, \bibinfo{author}{B.~Guilhot},
  \bibinfo{author}{M.~Beauvy}, \bibinfo{author}{M.~Benabdesselam},
  \bibinfo{author}{I.~Monnet}, \bibinfo{journal}{J. Nucl. Mater.}
  \bibinfo{volume}{385} (\bibinfo{year}{2009}) \bibinfo{pages}{304--307}.
%Type = Article
\bibitem[{Liu et~al.(2010)Liu, Le~Flem, B{\'e}chade, and
  Monnet}]{2010liu149JNM}
\bibinfo{author}{X.~Liu}, \bibinfo{author}{M.~Le~Flem},
  \bibinfo{author}{J.~B{\'e}chade}, \bibinfo{author}{I.~Monnet},
  \bibinfo{journal}{J. Nucl. Mater.} \bibinfo{volume}{401}
  (\bibinfo{year}{2010}) \bibinfo{pages}{149--153}.
%Type = Article
\bibitem[{Whittle et~al.(2010)Whittle, Blackford, Aughterson, Moricca, Lumpkin,
  Riley, and Zaluzec}]{2010whittle4362AM}
\bibinfo{author}{K.~Whittle}, \bibinfo{author}{M.~Blackford},
  \bibinfo{author}{R.~Aughterson}, \bibinfo{author}{S.~Moricca},
  \bibinfo{author}{G.~Lumpkin}, \bibinfo{author}{D.~Riley},
  \bibinfo{author}{N.~Zaluzec}, \bibinfo{journal}{Acta Mater.}
  \bibinfo{volume}{58} (\bibinfo{year}{2010}) \bibinfo{pages}{4362 -- 4368}.
%Type = Article
\bibitem[{Napp¨¦ et~al.(2011{\natexlab{a}})Napp¨¦, Maurice, Grosseau, Audubert,
  Thom¨¦, Guilhot, Beauvy, and Benabdesselam}]{2011nappe1503}
\bibinfo{author}{J.~Napp¨¦}, \bibinfo{author}{C.~Maurice},
  \bibinfo{author}{P.~Grosseau}, \bibinfo{author}{F.~Audubert},
  \bibinfo{author}{L.~Thom¨¦}, \bibinfo{author}{B.~Guilhot},
  \bibinfo{author}{M.~Beauvy}, \bibinfo{author}{M.~Benabdesselam},
  \bibinfo{journal}{J. Eur. Ceram. Soc.} \bibinfo{volume}{31}
  (\bibinfo{year}{2011}{\natexlab{a}}) \bibinfo{pages}{1503 -- 1511}.
%Type = Article
\bibitem[{Napp¨¦ et~al.(2011{\natexlab{b}})Napp¨¦, Monnet, Grosseau, Audubert,
  Guilhot, Beauvy, Benabdesselam, and Thom¨¦}]{2011nappe53JNM}
\bibinfo{author}{J.~Napp¨¦}, \bibinfo{author}{I.~Monnet},
  \bibinfo{author}{P.~Grosseau}, \bibinfo{author}{F.~Audubert},
  \bibinfo{author}{B.~Guilhot}, \bibinfo{author}{M.~Beauvy},
  \bibinfo{author}{M.~Benabdesselam}, \bibinfo{author}{L.~Thom¨¦},
  \bibinfo{journal}{J. Nucl. Mater.} \bibinfo{volume}{409}
  (\bibinfo{year}{2011}{\natexlab{b}}) \bibinfo{pages}{53 -- 61}.
%Type = Article
\bibitem[{Condon and Schober(1993)}]{1993condon1JNM}
\bibinfo{author}{J.~Condon}, \bibinfo{author}{T.~Schober}, \bibinfo{journal}{J.
  Nucl. Mater.} \bibinfo{volume}{207} (\bibinfo{year}{1993}) \bibinfo{pages}{1
  -- 24}.
%Type = Article
\bibitem[{Nordlander et~al.(1989)Nordlander, No, Besenbacher, and
  Myers}]{1989nordlander1990PRB}
\bibinfo{author}{P.~Nordlander}, \bibinfo{author}{J.~No},
  \bibinfo{author}{F.~Besenbacher}, \bibinfo{author}{S.~Myers},
  \bibinfo{journal}{Phys. Rev. B} \bibinfo{volume}{40} (\bibinfo{year}{1989})
  \bibinfo{pages}{1990}.
%Type = Article
\bibitem[{Arkhipov et~al.(2007)Arkhipov, Kanashenko, Sharapov, Zalavutdinov,
  and Gorodetsky}]{2007arkhipov1168JNM}
\bibinfo{author}{I.~Arkhipov}, \bibinfo{author}{S.~Kanashenko},
  \bibinfo{author}{V.~Sharapov}, \bibinfo{author}{R.~K. Zalavutdinov},
  \bibinfo{author}{A.~Gorodetsky}, \bibinfo{journal}{J. Nucl. Mater.}
  \bibinfo{volume}{363} (\bibinfo{year}{2007}) \bibinfo{pages}{1168--1172}.
%Type = Article
\bibitem[{Zhou et~al.(2001)Zhou, Sun, Wang, and Chen}]{2001zhou10001}
\bibinfo{author}{Y.~Zhou}, \bibinfo{author}{Z.~Sun}, \bibinfo{author}{X.~Wang},
  \bibinfo{author}{S.~Chen}, \bibinfo{journal}{J. Phys.: Condens. Matter}
  \bibinfo{volume}{13} (\bibinfo{year}{2001}) \bibinfo{pages}{10001}.
%Type = Article
\bibitem[{Kresse and Hafner(1993)}]{1993kresse558PRB}
\bibinfo{author}{G.~Kresse}, \bibinfo{author}{J.~Hafner},
  \bibinfo{journal}{PPhys. Rev. B} \bibinfo{volume}{47} (\bibinfo{year}{1993})
  \bibinfo{pages}{558}.
%Type = Article
\bibitem[{Kresse and Furthm¨¹ller(1996)}]{1996kressee11169PRB}
\bibinfo{author}{G.~Kresse}, \bibinfo{author}{J.~Furthm¨¹ller},
  \bibinfo{journal}{Phys. Rev. B} \bibinfo{volume}{54} (\bibinfo{year}{1996})
  \bibinfo{pages}{11169}.
%Type = Article
\bibitem[{Bl{\"o}chl(1994)}]{1994blochl17953PRB}
\bibinfo{author}{P.~E. Bl{\"o}chl}, \bibinfo{journal}{Phys. Rev. B}
  \bibinfo{volume}{50} (\bibinfo{year}{1994}) \bibinfo{pages}{17953}.
%Type = Article
\bibitem[{Perdew et~al.(1992)Perdew, Chevary, Vosko, Jackson, Pederson, Singh,
  and Fiolhais}]{1992perdew6671PRB}
\bibinfo{author}{J.~P. Perdew}, \bibinfo{author}{J.~A. Chevary},
  \bibinfo{author}{S.~H. Vosko}, \bibinfo{author}{K.~A. Jackson},
  \bibinfo{author}{M.~R. Pederson}, \bibinfo{author}{D.~J. Singh},
  \bibinfo{author}{C.~Fiolhais}, \bibinfo{journal}{Phys. Rev. B}
  \bibinfo{volume}{46} (\bibinfo{year}{1992}) \bibinfo{pages}{6671--6687}.
  \bibinfo{note}{PRB}.
%Type = Article
\bibitem[{Chadi(1977)}]{1977chadi1746PRB}
\bibinfo{author}{D.~Chadi}, \bibinfo{journal}{Phys. Rev. B}
  \bibinfo{volume}{16} (\bibinfo{year}{1977}) \bibinfo{pages}{1746}.
%Type = Article
\bibitem[{Sun and Zhou(1999)}]{1999sun1441PRB}
\bibinfo{author}{Z.~Sun}, \bibinfo{author}{Y.~Zhou}, \bibinfo{journal}{Phys.
  Rev. B} \bibinfo{volume}{60} (\bibinfo{year}{1999}) \bibinfo{pages}{1441}.
%Type = Article
\bibitem[{N{\o}rskov et~al.(1982)N{\o}rskov, Besenbacher, B{\o}ttiger, Nielsen,
  and Pisarev}]{1982Norskov1420PRL}
\bibinfo{author}{J.~N{\o}rskov}, \bibinfo{author}{F.~Besenbacher},
  \bibinfo{author}{J.~B{\o}ttiger}, \bibinfo{author}{B.~B. Nielsen},
  \bibinfo{author}{A.~Pisarev}, \bibinfo{journal}{Phys. Rev. Lett.}
  \bibinfo{volume}{49} (\bibinfo{year}{1982}) \bibinfo{pages}{1420--1423}.
%Type = Article
\bibitem[{Besenbacher et~al.(1985)Besenbacher, N{\o}rskov, Puska, and
  Holloway}]{1985besenbacher852PRL}
\bibinfo{author}{F.~Besenbacher}, \bibinfo{author}{J.~N{\o}rskov},
  \bibinfo{author}{M.~Puska}, \bibinfo{author}{S.~Holloway},
  \bibinfo{journal}{Phys. rev. lett.} \bibinfo{volume}{55}
  (\bibinfo{year}{1985}) \bibinfo{pages}{852}.
%Type = Article
\bibitem[{Jia et~al.(2012)Jia, Wang, Ou, Shi, and Ding}]{2012jia23ML}
\bibinfo{author}{L.~Jia}, \bibinfo{author}{Y.~Wang}, \bibinfo{author}{X.~Ou},
  \bibinfo{author}{L.~Shi}, \bibinfo{author}{W.~Ding}, \bibinfo{journal}{Mater.
  Lett.} \bibinfo{volume}{83} (\bibinfo{year}{2012}) \bibinfo{pages}{23--26}.
%Type = Article
\bibitem[{Liu et~al.(2009)Liu, Zhang, Zhou, Lu, Liu, and
  Luo}]{2009liu172103PRB}
\bibinfo{author}{Y.~L. Liu}, \bibinfo{author}{Y.~Zhang}, \bibinfo{author}{H.~B.
  Zhou}, \bibinfo{author}{G.~H. Lu}, \bibinfo{author}{F.~Liu},
  \bibinfo{author}{G.~N. Luo}, \bibinfo{journal}{Phys. Rev. B}
  \bibinfo{volume}{79} (\bibinfo{year}{2009}) \bibinfo{pages}{172103}.
%Type = Article
\bibitem[{Zhou et~al.(2010{\natexlab{a}})Zhou, Liu, Jin, Zhang, Luo, and
  Lu}]{2010zhou115010NF}
\bibinfo{author}{H.~B. Zhou}, \bibinfo{author}{Y.~L. Liu},
  \bibinfo{author}{S.~Jin}, \bibinfo{author}{Y.~Zhang}, \bibinfo{author}{G.~N.
  Luo}, \bibinfo{author}{G.~H. Lu}, \bibinfo{journal}{Nucl. Fusion}
  \bibinfo{volume}{50} (\bibinfo{year}{2010}{\natexlab{a}})
  \bibinfo{pages}{115010}.
%Type = Article
\bibitem[{Zhou et~al.(2010{\natexlab{b}})Zhou, Liu, Jin, Zhang, Luo, and
  Lu}]{2010zhoui025016NF}
\bibinfo{author}{H.~B. Zhou}, \bibinfo{author}{Y.~L. Liu},
  \bibinfo{author}{S.~Jin}, \bibinfo{author}{Y.~Zhang}, \bibinfo{author}{G.~N.
  Luo}, \bibinfo{author}{G.~H. Lu}, \bibinfo{journal}{Nucl. Fusion}
  \bibinfo{volume}{50} (\bibinfo{year}{2010}{\natexlab{b}})
  \bibinfo{pages}{025016}.
%Type = Article
\bibitem[{Zhang et~al.(2012)Zhang, Zhao, and Wen}]{2012zhang216JNM}
\bibinfo{author}{P.~Zhang}, \bibinfo{author}{J.~Zhao},
  \bibinfo{author}{B.~Wen}, \bibinfo{journal}{J. Nucl. Mater.}
  \bibinfo{volume}{429} (\bibinfo{year}{2012}) \bibinfo{pages}{216--220}.
%Type = Article
\bibitem[{Alkhamees et~al.(2013)Alkhamees, Zhou, Liu, Jin, Zhang, and
  Lu}]{20136alkhamees6JNM}
\bibinfo{author}{A.~Alkhamees}, \bibinfo{author}{H.~B. Zhou},
  \bibinfo{author}{Y.~L. Liu}, \bibinfo{author}{S.~Jin},
  \bibinfo{author}{Y.~Zhang}, \bibinfo{author}{G.~H. Lu}, \bibinfo{journal}{J.
  Nucl. Mater.} \bibinfo{volume}{437} (\bibinfo{year}{2013}) \bibinfo{pages}{6
  -- 10}.
%Type = Article
\bibitem[{Sun et~al.(2013)Sun, Jin, Li, Zhang, and Lu}]{2013Sun395JNM}
\bibinfo{author}{L.~Sun}, \bibinfo{author}{S.~Jin}, \bibinfo{author}{X.~C. Li},
  \bibinfo{author}{Y.~Zhang}, \bibinfo{author}{G.~H. Lu}, \bibinfo{journal}{J.
  Nucl. Mater.} \bibinfo{volume}{434} (\bibinfo{year}{2013})
  \bibinfo{pages}{395 -- 401}.
%Type = Article
\bibitem[{Puska et~al.(1981)Puska, Nieminen, and Manninen}]{1981puska3037PRB}
\bibinfo{author}{M.~Puska}, \bibinfo{author}{R.~Nieminen},
  \bibinfo{author}{M.~Manninen}, \bibinfo{journal}{Phys. Rev. B}
  \bibinfo{volume}{24} (\bibinfo{year}{1981}) \bibinfo{pages}{3037}.
%Type = Article
\bibitem[{Besenbacher et~al.(1985)Besenbacher, Myers, and
  N{\o}rskov}]{1985besenbacher55}
\bibinfo{author}{F.~Besenbacher}, \bibinfo{author}{S.~Myers},
  \bibinfo{author}{J.~N{\o}rskov}, \bibinfo{journal}{Nucl. Instrum. Methods
  Phys. Res. B} \bibinfo{volume}{7} (\bibinfo{year}{1985})
  \bibinfo{pages}{55--66}.
%Type = Article
\bibitem[{N{\o}rskov and Besenbacher(1987)}]{1987norskov475}
\bibinfo{author}{J.~N{\o}rskov}, \bibinfo{author}{F.~Besenbacher},
  \bibinfo{journal}{J. Less Common Met.} \bibinfo{volume}{130}
  (\bibinfo{year}{1987}) \bibinfo{pages}{475--490}.
%Type = Article
\bibitem[{Fu et~al.(2004)Fu, Willaime, and Ordej¨®n}]{2004fu175503PRL}
\bibinfo{author}{C.~C. Fu}, \bibinfo{author}{F.~Willaime},
  \bibinfo{author}{P.~Ordej¨®n}, \bibinfo{journal}{Phys. Rev. Lett.}
  \bibinfo{volume}{92} (\bibinfo{year}{2004}) \bibinfo{pages}{175503}.
%Type = Article
\bibitem[{Xu and Zhao(2009)}]{2009xu3170}
\bibinfo{author}{J.~Xu}, \bibinfo{author}{J.~Zhao}, \bibinfo{journal}{Nucl.
  Instrum. Methods Phys. Res. B} \bibinfo{volume}{267} (\bibinfo{year}{2009})
  \bibinfo{pages}{3170--3174}.
%Type = Article
\bibitem[{Duan et~al.(2010)Duan, Liu, Zhou, Zhang, Jin, Lu, and
  Luo}]{2010duan109JNM}
\bibinfo{author}{C.~Duan}, \bibinfo{author}{Y.~L. Liu}, \bibinfo{author}{H.~B.
  Zhou}, \bibinfo{author}{Y.~Zhang}, \bibinfo{author}{S.~Jin},
  \bibinfo{author}{G.~H. Lu}, \bibinfo{author}{G.~N. Luo}, \bibinfo{journal}{J.
  Nucl. Mater.} \bibinfo{volume}{404} (\bibinfo{year}{2010})
  \bibinfo{pages}{109--115}.
%Type = Article
\bibitem[{Ismer et~al.(2009)Ismer, Park, Janotti, and Van~de
  Walle}]{2009ismer184110PRB}
\bibinfo{author}{L.~Ismer}, \bibinfo{author}{M.~S. Park},
  \bibinfo{author}{A.~Janotti}, \bibinfo{author}{C.~G. Van~de Walle},
  \bibinfo{journal}{Phys. Rev. B} \bibinfo{volume}{80} (\bibinfo{year}{2009})
  \bibinfo{pages}{184110}.

\end{thebibliography}
\normalsize





% \end{thebibliography}
\newpage

Figure legend
%%line spacing used together with package package{setspace}
%\begin{spacing}{2.0}
%\end{spacing}

\begin{enumerate}

\item (Color online) Crystal structure of Ti$_3$SiC$_2$. Interstitial configurations of I-SiTi, I-Ti and I-SiC are also depicted by the frames of dashed lines. Bule, dark yellow, dark gray and dark cyan circles present H, Ti, C and Si atoms, respectively.\label{fig1}

\item (Color online) Trapping energy for H atom as a function of the number of H atom trapped by the Si vacancy in Ti$_3$SiC$_2$.\label{fig2}

\item (Color online) Atomic configuration and the corresponding isosurface of optimal charge density of H for different number of embedded H atoms at the Si vacancy in Ti$_3$SiC$_2$.\label{fig3}

\item (Color online) Isosurface of optimal charge density for H atom at I-SiC in Ti$_3$SiC$_2$.\label{fig4}

\item (Color online) Diffusion paths for H in Ti$_3$SiC$_2$. (a) In perfect Ti$_3$SiC$_2$. Site 1 represents the I-SiC sit, Site2 represents the nearest neighboring I-SiC sit. Site 3 represents the equivalent sit of Sit 1. (b) In Ti$_3$SiC$_2$ with a Si vacancy. The red ball represents a Si vacancy. Sit 1 represents the stable sit of an H in avacancy,Sites 3 represents the I-SiC sit. Site 4 represents the equivalent sit of Sit 3.\label{fig5}

\item (Color online) Diffusion energy profile for H in Ti$_3$SiC$_2$ with and without a Si vacancy. The corresponding diffusion paths are depicted in Fig. \ref{fig5}.\label{fig6}

\end{enumerate}

%\noindent{\large{Figure legend}}
\newpage

 \begin{figure}
 \includegraphics[width=8.5cm]{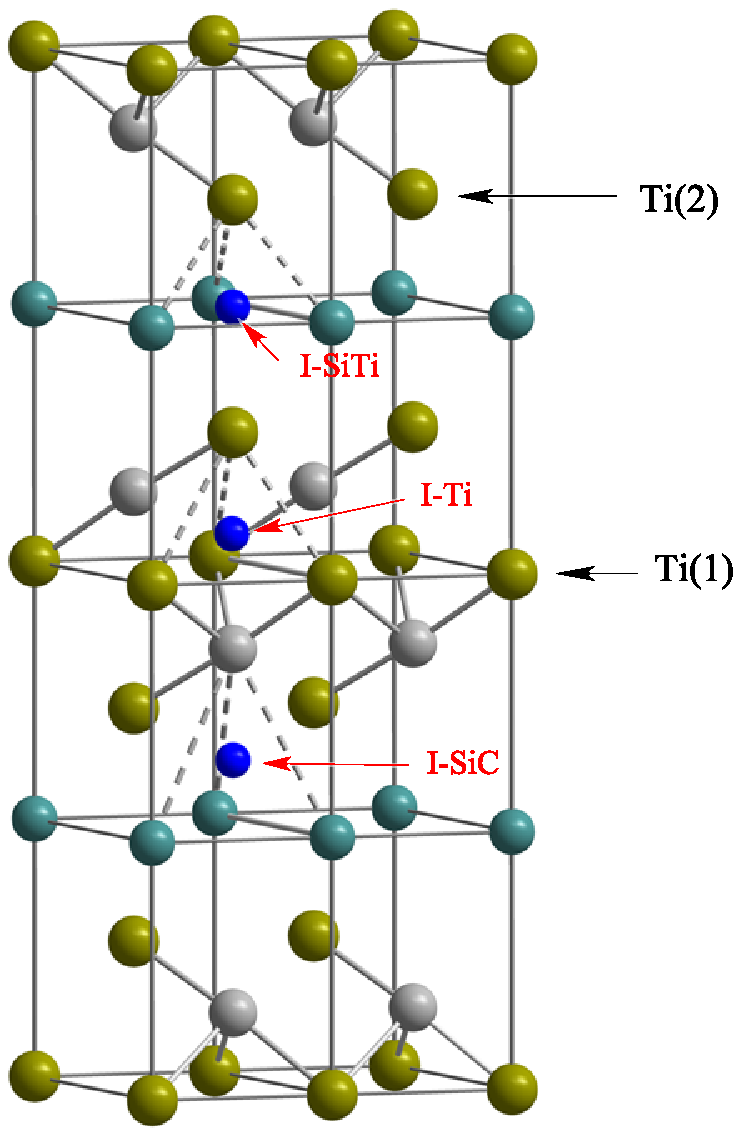}%
 \centering
 \caption{\label{fig1}}
 \end{figure}

 \begin{figure}
 \includegraphics[width=8.5cm]{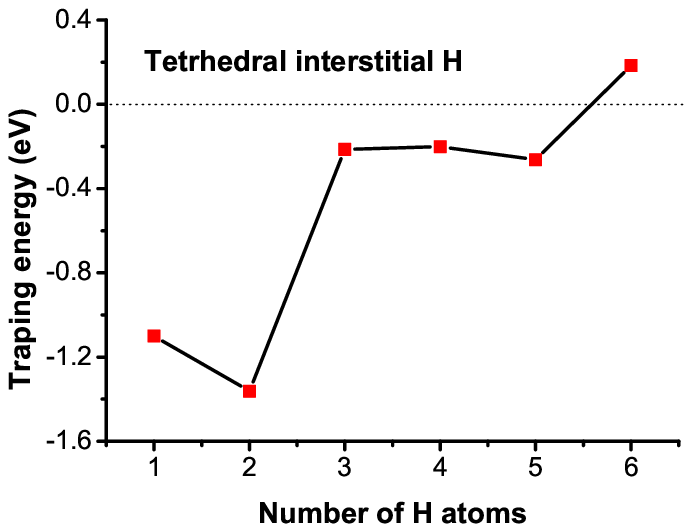}%
 \centering %
 \caption{\label{fig2}}
 \end{figure}

  \begin{figure}
  \includegraphics[width=8.5cm]{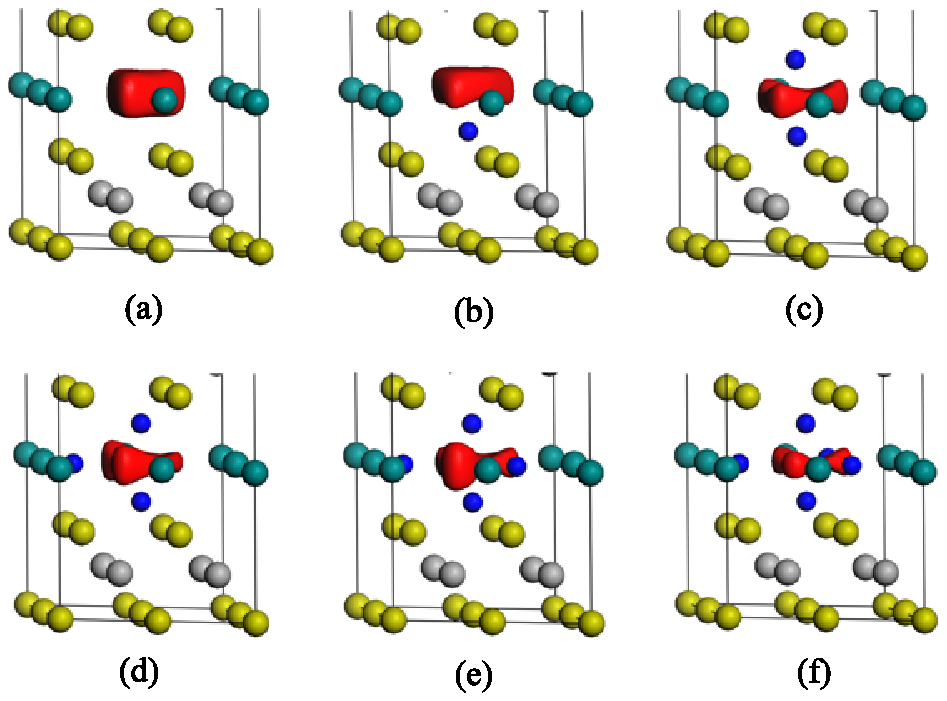}%
  \centering
  \caption{\label{fig3}}
  \end{figure}

 \begin{figure}
 \includegraphics[width=8.5cm]{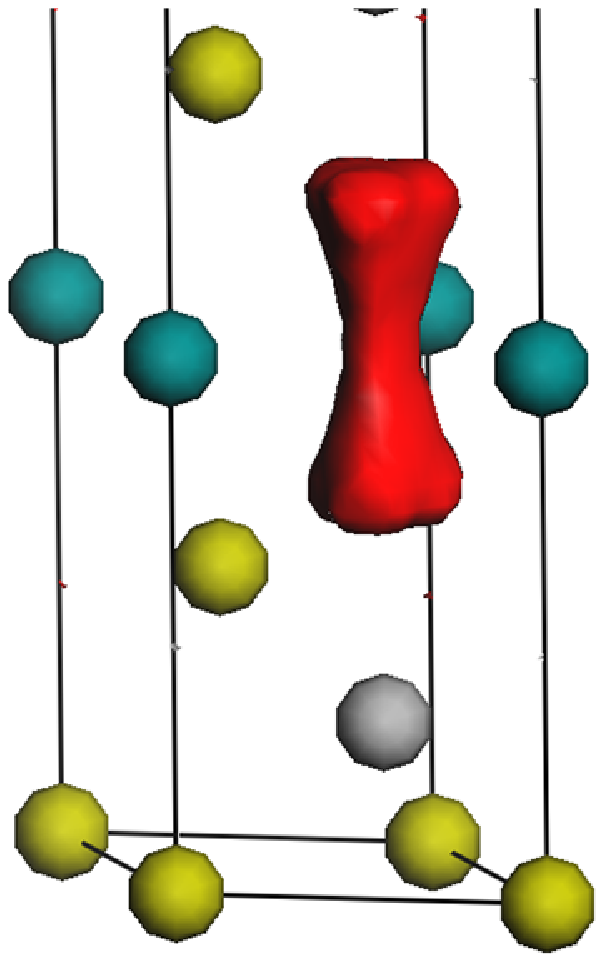}%
 \centering
 \caption{\label{fig4}}
 \end{figure}

 \begin{figure}
 \includegraphics[width=8.5cm]{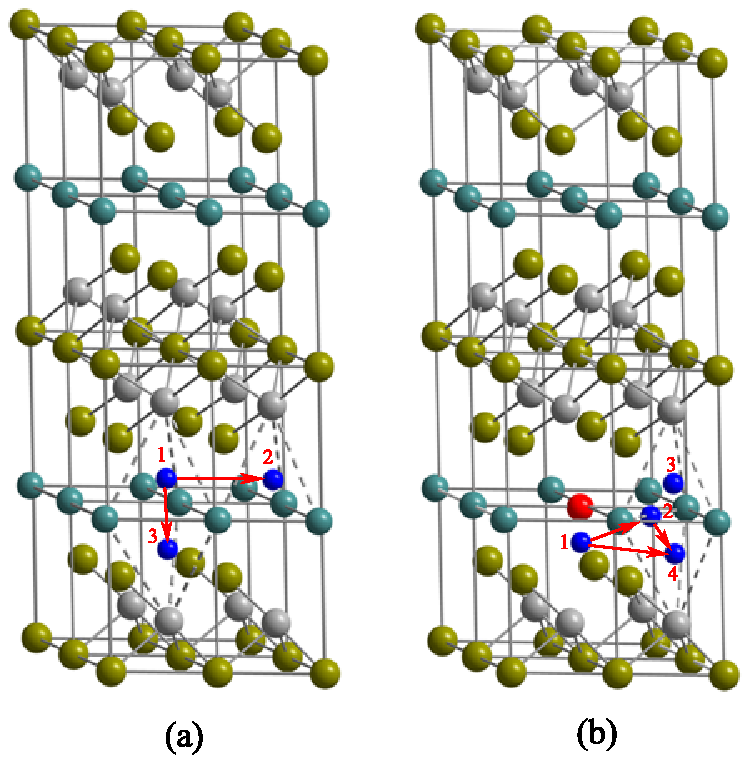}%
 \centering
 \caption{\label{fig5}}
 \end{figure}

  \begin{figure}
 \includegraphics[width=8.5cm]{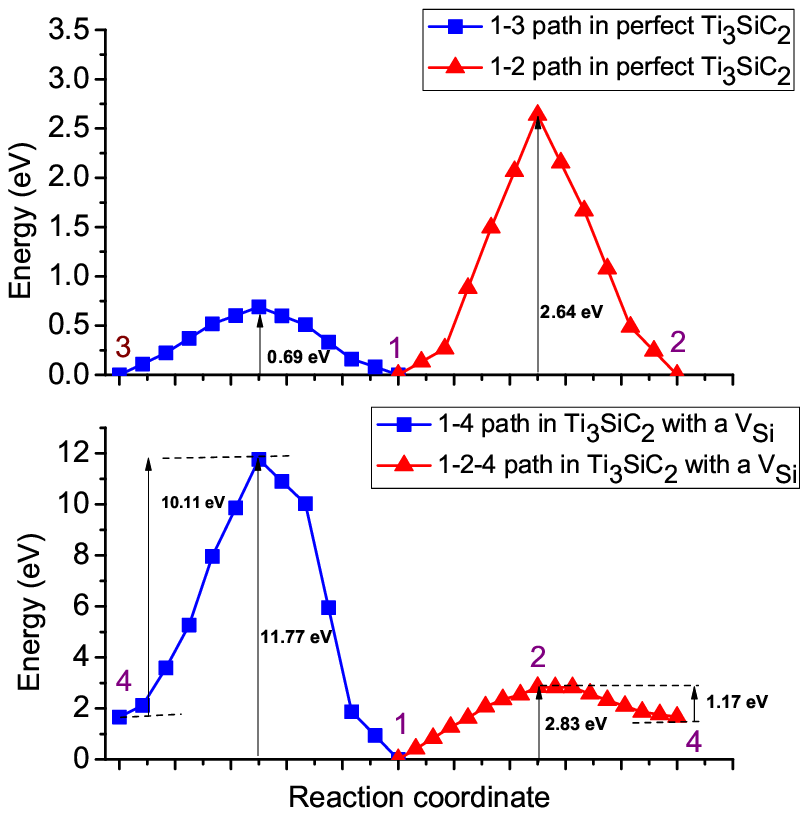}%
 \centering
 \caption{\label{fig6}}
 \end{figure}

\end{document}